\documentclass[prl,twocolumn,showpacs]{revtex4}
\usepackage{graphicx,epsf}
\usepackage{bm, bbm}      % bold math

\newcommand{\bone}{\mathbbm{1}}

\newcommand{\Pib}{\mbox{\boldmath $\Pi $}}

\newcommand{\rhobar}{\bar{\rho}}

\newcommand{\Sbar}{\bar{S}}

\newcommand{\bp}{{\bf p}}
\newcommand{\bq}{{\bf q}}
\newcommand{\bk}{{\bf k}}
\newcommand{\br}{{\bf r}}
\newcommand{\be}{{\bf e}}

\newcommand{\bR}{{\bf R}}
\newcommand{\bB}{{\bf B}}
\newcommand{\bA}{{\bf A}}

\newcommand{\mtilde}{\tilde{m}}

\newcommand{\beq}{\begin{equation}}
\newcommand{\beqn}{\begin{eqnarray}}
\newcommand{\eeq}{\end{equation}}
\newcommand{\eeqn}{\end{eqnarray}}
\newcommand{\nn}{\nonumber}

\newcommand{\da}{\downarrow}
\newcommand{\ua}{\uparrow}

\newcommand{\Bmath}{\mathcal{B}}
\newcommand{\Amath}{\mathcal{A}}

\begin{document}

\def\tende#1{\,\vtop{\ialign{##\crcr\rightarrowfill\crcr
\noalign{\kern-1pt\nointerlineskip}
\hskip3.pt${\scriptstyle #1}$\hskip3.pt\crcr}}\,}

\title{From Fractional Chern Insulators to a Fractional Quantum Spin Hall Effect}
\author{M. O. Goerbig}

\affiliation{
Laboratoire de Physique des Solides, CNRS UMR 8502, Univ. Paris-Sud, F-91405 Orsay cedex, France}

\begin{abstract}

We investigate the algebraic structure of flat energy bands a partial filling of which may give rise to a fractional quantum anomalous 
Hall effect (or a fractional Chern insulator) and a fractional quantum spin Hall effect. Both effects arise in the case of a 
sufficiently flat energy band as well as a roughly flat and homogeneous Berry curvature, such that the global Chern number, 
which is a topological
invariant, may be associated with a local non-commutative geometry. This geometry is similar to the more familiar situation of 
the fractional quantum Hall effect in two-dimensional electron systems in a strong magnetic field. 

\end{abstract}
\maketitle

\section{Introduction}

One of the most fascinating effects that arise from the interplay between topology and strong electron-electron interactions is
certainly the fractional quantum Hall effect (FQHE) \cite{TSG}.
Once a single Landau level, that may be identified with an infinitely flat band
of two-dimensional (2D) electrons in a strong magnetic field, is only partially filled, the electronic interactions remain as the 
only relevant energy scale, which causes for certain fillings the formation of incompressible quantum liquids with fractionally charged 
quasi-particles \cite{laughlin}.

The topological properties underlying the FQHE are encoded in the Chern number, which is non-zero for a Landau level \cite{TKNN} and
that is a global manifestation of the non-commutative geometry of quantum states in a single level. 
Indeed, when the electron dynamics is restricted to a single Landau level, 
the position operator $\br_j$ of the $j$-th electron (of charge $-e$) is replaced by that of the 
centre of the cyclotron motion (``guiding centre''), 
$\bR_j=\br_j-\Pib_j\times \be_z/eB$, in terms of the kinetic momentum $\Pib_j=\bp_j+e\bA(\br_j)$,
where the vector potential $\bA(\br)$ yields the magnetic field $\bB=\nabla\times\bA(\br)=B\be_z$, and $\bp_j=-i\hbar\nabla_j$
is the canonical momentum of the $j$-th electron. Canonical quantisation then yields the commutation relations 
\beq\label{eq:commGC}
[X_j,Y_{j'}]=il_B^2\delta_{j,j'}
\eeq
for the components of the guiding-centre operator $\bR_j=(X_j,Y_j)$, in terms of the magnetic length 
$l_B=\sqrt{\hbar/eB}\simeq 26\,\text{nm}/\sqrt{B\text{[T]}}$.

These properties of the quantum Hall effect have been revisited in the framework of 2D topological insulators \cite{TIrev}.
Indeed, in an early piece of
work, Haldane pointed out that an integer quantum Hall effect may arise in the absence of a magnetic field if an energy band acquires
a non-zero Chern number $C$ \cite{haldane88}. This yields a non-zero Hall conductance, $\sigma_{xy}=Ce^2/h$, and the effect is known
as the quantum anomalous Hall effect. In order to obtain a non-zero Chern number, time-reversal symmetry must be broken -- indeed, 
Haldane's proposal consists of an inhomogeneous distribution of a (zero net) flux in the unit cell of a honeycomb lattice 
\cite{haldane88} that may though be difficult to achieve in an experimental situation. More recently, Kane and Mele have proposed a novel 
class of topological insulators that respect time-reversal symmetry and that may be viewed as two (spin-dependent) copies of Haldane's
original model \cite{KM05}. Although the total Hall conductance is then zero, the difference 
$\Delta\sigma_{xy}=\sigma_{xy,\ua}-\sigma_{xy,\da}$ happens to be quantised (quantum spin Hall effect). This situation may 
in principle arise in graphene with spin-orbit coupling -- however, practically the intrinsic spin-orbit coupling in graphene turns
out to be too small to yield a measurable effect. More successful was the prediction of a quantum spin Hall effect in HgTe/CdTe 
heterostructures by Bernevig, Hughes and Zhang \cite{BHZ}, which has been confirmed experimentally \cite{molenkamp}.

Based on these findings, it is natural to speculate whether the correspondance between the (integer) quantum Hall effect and
topological insulators (quantum anomalous Hall effect and quantum spin Hall effect) remains valid also in the strongly-correlated
limit, that is whether there is an effect equivalent 
to the FQHE in topological insulators. Very recently, flat bands with a non-zero Chern
number have been modeled theoretically in various kinds of 2D lattices \cite{Tang,Neupert1,Sun,Fiete}. 
All models involve time-reversal symmetry breaking and a fine tuning of the lattice hopping parameters
and may be viewed as generalisations of Haldane's original idea \cite{haldane88}. 
First numerical indications that partially filled flat bands with non-zero Chern number may support a fractional quantum anomalous
Hall effect (FQAHE) 
\cite{Neupert1,Sheng} have been confirmed by Regnault and Bernevig \cite{RB}, who obtained convincing evidence that the 1/3-filled
band is described by a Laughlin state with the usual fractionally charged quasi-particles. In addition to the construction of
explicit wave functions for partially filled bands \cite{Qi}, the analytic work by Parameswaran, Roy, and Sondhi \cite{PRS} 
aimed at a connection between the \textit{global} properties of a Chern band, that is a non-zero Chern number, and a more \textit{local} 
description, in terms of non-commutative geometry (\ref{eq:commGC}).

Whereas the approach by Parameswaran, Roy, and Sondhi \cite{PRS} is based on a small-wave-vector expansion, the present
paper provides a theoretical framework that is \textit{a priori} valid within the whole Brillouin zone. However, also in this 
framework, the Girvin-MacDonald-Platzman (GMP) algebra \cite{GMP}, which arises naturally in Landau levels from the 
commutation relation (\ref{eq:commGC}), is retrieved only for sufficiently flat Berry curvatures. A criterion for the flatness
of the Berry curvature, as compared to the flat-energy-band limit, is provided with the help of a simplified two-band model.
Furthermore, we discuss a generalisation to time-reversal-symmetric models with flat bands and
strong spin-orbit coupling. Similarly to the non-interacting case, these models may be viewed as two copies of the FQAHE for each
of the Kramers pairs and they may yield a fractional quantum spin Hall effect (FQSHE) if electron-electron interactions are 
taken into account. This effect has been investigated in recent numerical studies \cite{Neupert2}. We find that the 
commutation relations (\ref{eq:commGC}) are spin-dependent, as well as the resulting GMP
or $W_{\infty}$ algebra \cite{GMP},
which governs the electronic density fluctuations in a single band. Implications for the wave functions and the pseudopotentials 
\cite{PP} are equally discussed.

\section{Fractional Chern Insulator}

We consider the one-particle states $\psi_{\bk,\lambda}$, described by a wave vector $\bk$ in the first Brillouin zone in a band
$\lambda$ that is sufficiently well separated in energy from the other bands. In this case, the low-energy electronic
properties are obtained from a projection of the electronic dynamics to this single band. The state may be written as a sum,
\beq\label{eq01}
\psi_{\bk,\lambda} = \sum_{a}\alpha_{\bk,a}u_{\bk,\lambda}^a,
\eeq
of Bloch states $u_{\bk,\lambda}^a$, for each of the sublattices labeled by the index $a$, where $\alpha_{\bk,a}$ is the amplitude of the
state on the sublattice $a$. When projected to a single band, the reciprocal-space displacement operator 
$\exp(i\bq\cdot\hat{\br})$, that is the one-particle density operator, needs to account 
for a contribution that stems from the Bloch functions
\beq\label{eq02}
e^{i\bq\cdot\hat{\br}}\rightarrow e^{i\bq\cdot\hat{\br}}T_{\lambda}(\bq),
\eeq
where 
\beq\label{eq03}
T_{\lambda}(\bq)\equiv \sum_{\bk}\sum_a u_{\bk+\bq,\lambda}^a u_{\bk,\lambda}^{a *}.
\eeq
If we consider an infinitesimal displacement $\delta\bq$, one may Taylor-expand the above expression to lowest order,
$u_{\bk+\delta \bq,\lambda}^a u_{\bk,\lambda}^{a *}\simeq 1 - i\delta\bq\cdot \Amath_{\bk,\lambda}$, where
$\Amath_{\bk,\lambda}=i\sum_{a}u_{\bk,\lambda}^{a *}\nabla_{\bk}u_{\bk,\lambda}^{a}$ is the Berry connection in the band $\lambda$.
One therefore notices that the generator of infinitesimal reciprocal-space displacements reads
\beq\label{eq04}
\hat{\bR}(\bk)=\hat{\br}-\Amath_{\bk,\lambda}
\eeq
when the dynamics is restricted to a single band. This generator is a local quantity in the sense that it depends on the 
reciprocal-space position $\bk$ and thus on the state that it acts on. 

In order to obtain an integral expression of the displacement operator for a wave vector $\bq$ that is not necessarily infinitesimal,
one may use 
\beq\label{eq05}
T_{\lambda}(\bq)=\lim_{N\rightarrow\infty}\left(1-i\frac{\bq}{N}\cdot\Amath_{\bk,\lambda}\right)^N\simeq 
e^{-i\int_{\bk}^{\bk+\bq}d\bk'\cdot\Amath_{\bk',\lambda}},
\eeq
which is the expression obtained by Parameswaran, Roy, and Sondhi \cite{PRS}. Notice that the last identity should be interpreted in 
a path-integral sense, where one may choose the direct path from $\bk$ to $\bk+\bq$. The projected density operator in the band $\lambda$
may finally be written as 
\beq\label{eq06}
\rhobar_{\lambda}(\bq)=\sum_{\bk}e^{-i\int_{\bk}^{\bk+\bq}d\bk'\cdot\Amath_{\bk',\lambda}}\psi_{\bk+\bq,\lambda}\psi_{\bk,\lambda}^*.
\eeq

\begin{figure}
\centering
\includegraphics[width=5.5cm,angle=0]{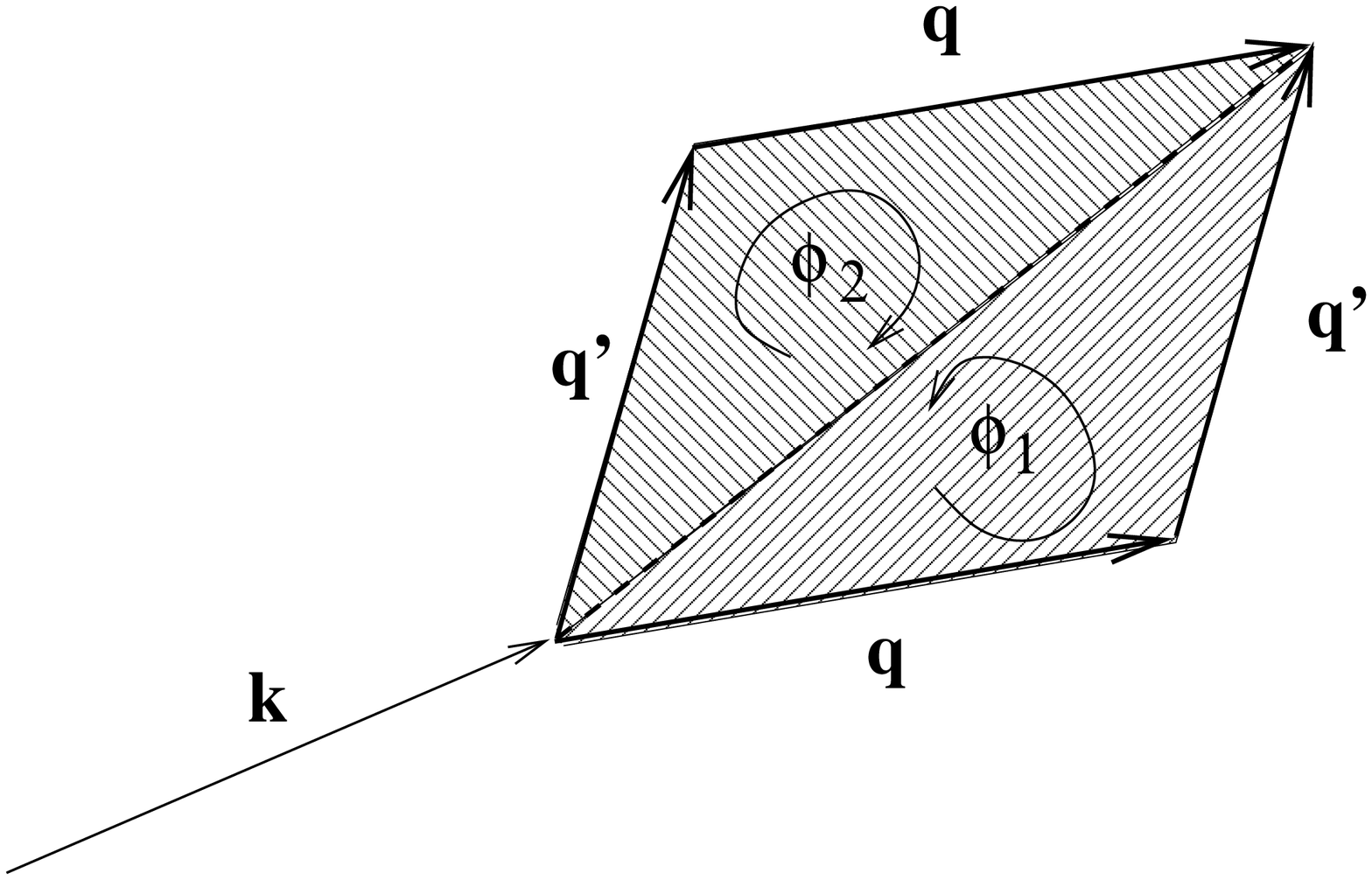}
\caption{\footnotesize{Flux $\phi_1=\phi(\bk,\bq,\bq')$ and $\phi_2=\phi(\bk,\bq',\bq)$ occuring in the expression (\ref{eq07}) 
for the commutation relations of the projected density operators.
}}
\label{fig01}
\end{figure}

This result has been derived in Ref. \cite{PRS} within a short-wave-vector expansion, but one notices that, by virtue
of Eq. (\ref{eq05}), it is valid within the entire Brillouin zone.
The additional phase in the expression (\ref{eq06}) for the projected density operator induces unusual commutation relations,
\beqn\label{eq07}
%\nn
\left[\rhobar_{\lambda}(\bq),\rhobar_{\lambda}(\bq')\right]&=&\sum_{\bk}\left[e^{-i\phi(\bk,\bq',\bq)}-e^{-i\phi(\bk,\bq,\bq')}\right]
\\
\nn
&&\times e^{-\int_{\bk}^{\bk+\bq}d\bk'\cdot\Amath_{\bk',\lambda}}\psi_{\bk+\bq+\bq',\lambda}\psi_{\bk,\lambda}^*,
\eeqn
where we have defined the phases 
$\phi(\bk,\bq,\bq')=(\int_{\bk}^{\bk+\bq}+\int_{\bk+\bq}^{\bk+\bq+\bq'}+\int_{\bk+\bq+\bq'}^{\bk})d\bk'\cdot\Amath_{\bk',\lambda}$ 
as the flux in the triangle with the points $\bk$, $\bk+\bq$, and $\bk+\bq+\bq'$
(see Fig. \ref{fig01}).

\subsection{Flat Berry curvature}

Equation (\ref{eq07}) may be viewed as a preliminary non-local (in reciprocal space) form of the GMP or $W_{\infty}$ algebra, and it
does not yet provide us with the algebraic structure underlying the fractional quantum Hall effect known from Landau levels. Indeed, 
for the latter, the flux density is homogeneous and created by a homogeneous magnetic field in real space -- a similar commutation 
relation is then obtained from the non-commutativity of the guiding-centre coordinates (\ref{eq:commGC}), in which case 
$\rhobar_B(\bq)=\exp(i\bq\cdot\bR)$, such that one obtains the GMP algebra 
$[\rhobar_B(\bq),\rhobar_B(\bq')]=2i\sin(\bq\wedge\bq'l_B^2/2)\rhobar(\bq+\bq')$ \cite{GMP}, where $\bq\wedge\bq'=q_xq_y'-q_yq_x'$.
In the case of a Chern band, however, the Berry curvature is generally not homogeneous. 

In spite of this drawback, let us for the moment consider the theoretical limit of a homogeneous non-zero
Berry curvature, which was assumed
in Ref. \cite{PRS} and that must not be confounded with the limit of 
an infinitely flat energy band. This point is illustrated in a particular two-band example below. For a homogeneous
Berry curvature $\Bmath_{\lambda}=\nabla_{\bk}\times\Amath_{\bk,\lambda}$, 
the phases occuring in the expression (\ref{eq07}) for the commutation relations do no longer depend on the 
precise position of the triangle $\bk$, $\bk+\bq$, $\bk+\bq+\bq'$ in reciprocal space. The fluxes $\phi_1=\phi(\bk,\bq,\bq')$ 
and $\phi_2=\phi(\bk,\bq',\bq)$ are then simply given by the area of the triangles 
spanned by the vectors $\bq$ and $\bq'$ times the Berry curvature, 
$\phi_1=-\phi_2=\Bmath_{\lambda} \bq \wedge\bq'/2$, such that the commutation relations (\ref{eq07}) reduce to the GMP algebra
\beq\label{eq08}
\left[\rhobar_{\lambda}(\bq),\rhobar_{\lambda}(\bq')\right]= 
2i\sin\left(\frac{\bq\wedge\bq'}{2}\Bmath_{\lambda}\right)\rhobar_{\lambda}(\bq+\bq').
\eeq
In this expression, the Berry curvature, which is given in terms of the Chern number $C_{\lambda}$ of the band and the surface $A_{BZ}$
of the first Brillouin zone \cite{PRS}
\beq\label{eq09}
\Bmath_{\lambda}=\frac{2\pi C_{\lambda}}{A_{BZ}},
\eeq
plays the role of an effective \textit{magnetic length}, $l_B^2=\Bmath_{\lambda}\sim a_0^2$, where $a_0$ is the lattice 
spacing in the real
lattice. In contrast to the quantum Hall effect and physically accessible magnetic fields, this is a very small length scale such that
the corresponding putative magnetic fields $B^*\equiv \hbar A_{BZ}/2\pi e C_{\lambda}\sim h/e a_0^2$
involved in Chern insulators are exremely large (some $10^4$ T for lattice spacings on the order of $a_0\sim$ \AA).

In the discussed limit of a homogeneous Berry curvature, the effective Hamiltonian that describes the electronic dynamics in a single
Chern band finally reads, in second quantisation,
\beq\label{eq10}
H_{\lambda}=\sum_{\bk}\epsilon_{\lambda,\bk}c_{\lambda,\bk}^{\dagger}c_{\lambda,\bk} + \frac{1}{2}\sum_{\bk}v(\bk)
\rhobar_{\lambda}(-\bk)\rhobar_{\lambda}(\bk),
\eeq
where $c_{\lambda,\bk}^{(\dagger)}$ annihilates (creates) an electron in the Bloch state $\psi_{\bk,\lambda}$ and that is accompanied
by the commutation relation (\ref{eq08}). Furthermore,
$v(\bk)$ is the Fourier transformation of an interaction that is considered to depend only on the relative distance between 
electrons in the original lattice. The precise form of this interaction potential determines to what extent a FQAHE, that is sensitive
to its behaviour at short distances, can be stabilised. Indeed, for a nearest-neighbour repulsion in the checkerboard lattice, 
Regnault and Bernevig have found clear numerical evidence for a Laughlin state at $\nu=2\pi \Bmath_{\lambda} n_{el}=1/3$ \cite{RB}.

Whereas the precise form of the effective interaction potential between two electrons in different states in a Chern band remains 
an open issue, partial insight may be obtained from an approximation in which one neglects the particular topology of the first
Brillouin zone and considers the flat Berry curvature as a homogeneous magnetic field acting on electrons in a 2D plane. This view point
is somewhat reminiscent of that of coarse-grained positions $[X,Y]=-i\Bmath_{\lambda}$ invoked in Ref. \cite{PRS}. The latter is rather 
delicate in the sense that coarse graining requires a large length scale over which one averages such as to omit details on the lattice
length scale. However, in the present problem all length scales coincide because 
$l_B=\sqrt{\Bmath_{\lambda}}\sim a_0$, and the assumption requires
therefore a flat Berry curvature. A criterion for the flatness of the Berry curvature is discussed below. 

Once mapped from the first Brillouin zone to a non-compact 2D plane, the interaction potential relevant for the stability of FQAHE states
may be described in terms of Haldane's pseudo-potentials \cite{PP}
\beq\label{eqPP}
V_{\ell}=\frac{1}{2\pi}\sum_{\bk}v(\bk)L_{\ell}(q^2\Bmath)e^{-q^2\Bmath/2}
\eeq
which are the same as those for electrons in a single Landau level if one replaces the magnetic length by the Berry curvature $\Bmath$.
However, this approach, which relies on a description of single-particle wave functions in the 2D plane, needs to be handled with care,
and single-particle wave functions in the torus geometry seem more appropriate in view of the topology of the first Brillouin zone. 
Further analytical and numerical studies are required to elucidate the issue of the precise form of the interactions required to
stabilise FQAHE states in Chern bands.

The model (\ref{eq10}) and the analogy with the FQHE in 2D electron systems in a strong magnetic field 
allow one to investigate qualitatively the validity of the underlying assumptions that have been made so 
far. The \textit{flat-band} assumption, which amounts to setting the dispersion $\epsilon_{\lambda,\bk}$ constant,
corresponds to the limit of an infinitely flat Landau level in the absence of disorder. Even if this assumption is usually
not valid in a true experimental situation, it is reasonable in the description of the FQHE, which is protected by an energy gap
that is a substantial fraction of the energy scale $e^2/\epsilon l_B$, where $\epsilon$ is the dielectric constant of the host
material. The poor-man's replacement $l_B\rightarrow \Bmath$ in the pseudo-potential expansion (\ref{eqPP})
leads to the expectation that the energy scales, and thus the gap of the
Laughlin state, are much larger in fractional Chern insulators than in the usual FQHE \cite{Tang}. The flat-band assumption is therefore
expected to be valid as long as the band width of $\epsilon_{\lambda,\bk}$ is smaller than this energy scale. 

However, this flat-band assumption should not be confounded with the \textit{flat-curvature} assumption. Consider for instance a 
two-band model that is generically described by a Hamiltonian 
\beq\label{eq11}
H_{\text{2-band}}=d_0(\bk)\bone + \vec{d}(\bk)\cdot\vec{\sigma},
\eeq
where $\bone$ is the $2\times 2$ identity matrix and $\vec{\sigma}=(\sigma_x,\sigma_y,\sigma_z)$ in terms of the Pauli matrices. The 
dispersion of the two bands is given by $\epsilon_{\lambda,\bk}=d_0(\bk)+\lambda |\vec{d}(\bk)|$, where $\lambda=\pm$ for the two bands. 
The flat-band limit for the band $\lambda$ is therefore given by $d_0(\bk)\simeq -\lambda|\vec{d}(\bk)|$ for all values of the 
wave vector. One notices from this expression that, within the two-band model, the flat-band limit cannot be satisfied simultaneously
for $\lambda=+$ and $-$.
In contrast to the dispersion relation, the Berry curvature 
\beq\label{eq12}
\Bmath_{\bk}=\frac{1}{4\pi|\vec{d}(\bk)|^3}\vec{d}(\bk)\cdot\partial_{k_x}\vec{d}(\bk)\times\partial_{k_y}\vec{d}(\bk)
\eeq
is independent of the term $d_0(\bk)\bone$ in the two-band model (\ref{eq11}), where we have dropped the index $\lambda$ in the
expression. Therefore, although one may use $d_0(\bk)\bone$ to engineer a flat band, it is not sufficient to render the Berry
curvature flat, which is unaffected by the term.

Deviations from the flat Berry curvature may be identified, in the case of a FQHE in 2D electron systems in a strong magnetic field, with
an inhomogeneous distribution of the flux density and thus a variation of the local filling factor in real space. If this variation
is small and slow, the Laughlin state remains the ground state because of its incompressibility or else the presence of an energy 
gap. 
On the other hand, if there is a strong variation of the magnetic field or the electronic density, the incompressible quantum liquid 
breaks up into droplets of quantum liquids with different quantum numbers \cite{inhomFQHE}. 

\subsection{Criterion for the flatness of the Berry curvature}

In order to further investigate the flat-curvature assumption, which allows for the occurence of a FQAHE in Chern insulators, we 
use the two-band model (\ref{eq11}) and consider the non-zero Chern number $C_{\lambda}$ to arise from hidden Dirac points located
at some wave vectors $\bk_j$. For a topological phase transition from a Chern insulator to a trivial band insulator, the band gap must 
vanish at at least one of these wave vectors, and one obtains a true Dirac point at $\bk_j$ 
with a linear dispersion relation \cite{TIrev}.
In the vicinity of these particular wave vectors, the two-band model may be approximated by
\beq\label{eq13}
%\nn
d_{x/y}(\bk)=v_{x/y}(k_{x/y}-k_{x/y,j}), %\qquad d_y(\bk)=v_y(k_y-k_{y,j}), 
\qquad d_z(\bk)=\xi_j M_j,
\eeq
where we have omitted the term $d_0(\bk)\bone$, which is irrelevant for the discussion of the Berry curvature, and where we have set 
$\hbar=1$. The velocities $v_x$ and $v_y$ arise in the expansion of the two-band model (\ref{eq11}), and $\xi_j$ is the sign of 
the mass gap with strength $M_j>0$. In this case, the Berry 
curvature reads \cite{fuchs}
\beq\label{eq14}
\Bmath_{\bk}\simeq \frac{1}{4\pi}\frac{\xi_j M_j v_x v_y}{\left[M_j^2+v_x^2(k_x-k_{x,j})^2 + v_y^2(k_y-k_{y,j})^2\right]^{3/2}},
\eeq
and one notices that the mass term $M_j$ introduces a characteristic wave vector $\kappa_{x/y}=M_j/v_{x/y}$. For values 
$|k_{x/y}-k_{x/y,j}|\lesssim \kappa_{x/y}$, the Berry curvature may thus be considered as roughly constant
whereas for wave vectors larger than $\kappa_{x/y}$, the curvature drops to zero. This yields the criterion 
\beq\label{eq15}
\kappa_{x/y}=\frac{M_j}{v_{x/y}}\sim \frac{1}{a_0},
\eeq
in which case the Berry curvature may be considered as roughly homogeneous within the first Brillouin zone, which occupies a 
characteristic surface of $\sim 1/a_0^2$. Furthermore, one notices that in the vicinity of a topological phase transition, where
$M_j\rightarrow 0$, the characteristic wave vector becomes extremely small, and the approximation of a homogeneous Berry curvature
is therefore unjustified.

\subsection{Illustration in the checkerboard lattice}

\begin{figure}
\centering
\includegraphics[width=5.5cm,angle=0]{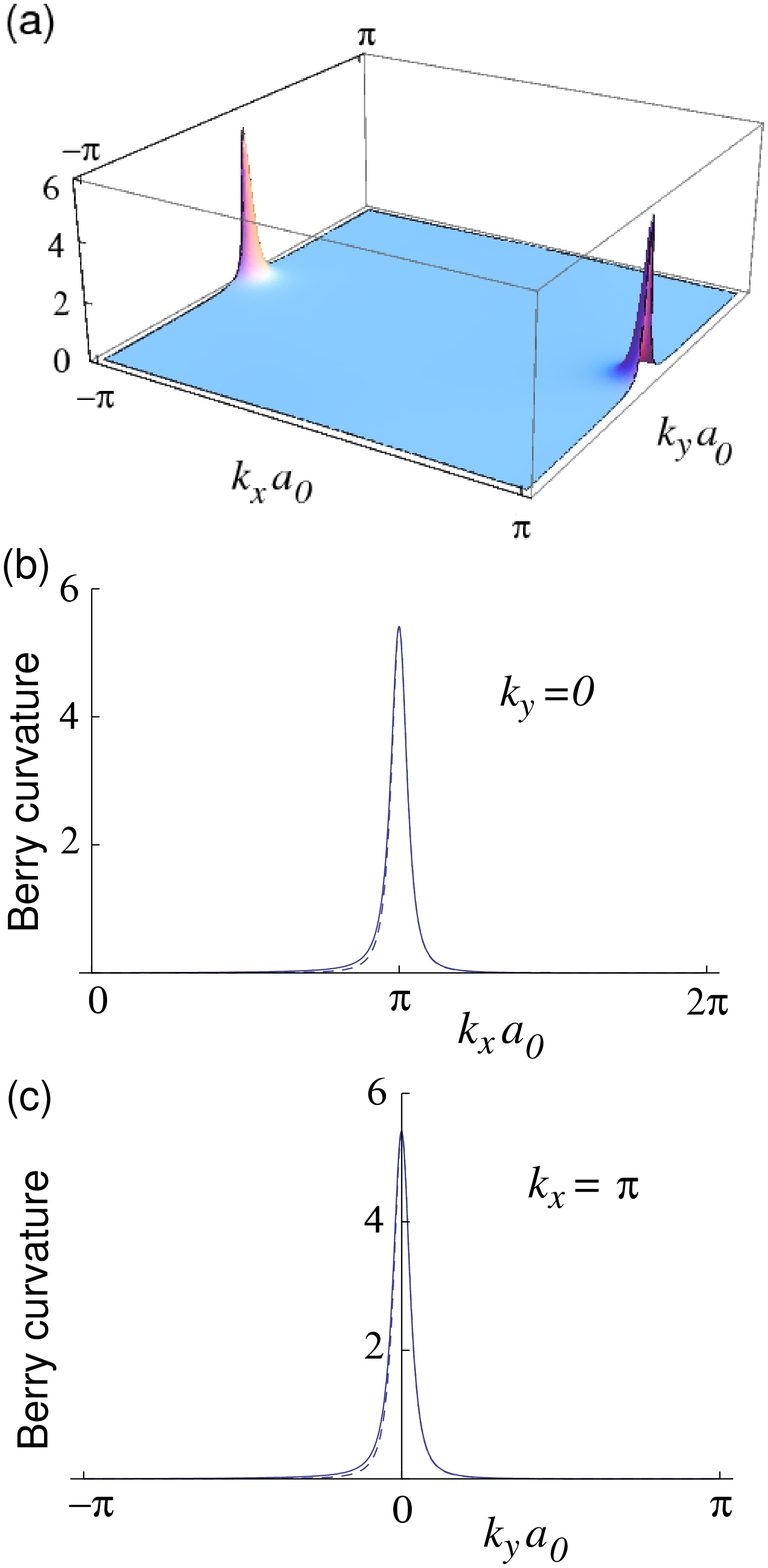}
\caption{\footnotesize{(a) Berry curvature of the checkerboard-lattice model with $t_2/t_1=(2-\sqrt{2})/2$, $\phi=\pi/4$, and $m/t_1=1$, 
over the whole Brillouin zone. (b) Berry curvature for $k_y=0$ (full line), in comparison to approximate formula (\ref{eq14}) (dashed
line). (c) Same for $k_x=\pi$.
}}
\label{fig02}
\end{figure}

One of the models used in the discussion of flat Chern bands is the checkerboard lattice, which consists of two interpenetrating
square lattices \cite{Sun,RB}.
In order to illustrate the above criterion of flat Berry curvature, 
we consider this lattice in the form proposed in Ref. \cite{RB}, with the 
off-diagonal term 
\beqn\nn
d_x(\bk)+id_y(\bk) &=& t_1 e^{i\phi}\left[1+e^{i(k_y - k_x)a_0}\right]\\
&&+t_1 e^{-i\phi}\left[e^{ik_ya_0}+e^{-ik_xa_0}\right]
\eeqn
and the diagonal term
\beq
d_z(\bk)=2t_2\left[\cos(k_xa_0)-\cos(k_ya_0)\right] + m
\eeq
in the Hamiltonian (\ref{eq11}). Whereas $t_1$ describes nearest-neighbour hopping, $t_2$ denotes the anisotropy in the 
next-nearest-neighbour hopping between the two sublattices -- for more details on the model, we refer the reader to the 
literature \cite{Sun,RB}. 
The phase $\phi$ associated with nearest-neighbour hopping breaks time-reversal symmetry 
unless $\phi= n\pi$, in terms of an integer $n$, whereas the term $m$ describes lattice-inversion symmetry breaking. 
In the original model, a term $d_0(\bk)$ was taken into account to render one of the 
bands as flat as possible, in addition to the particular choice $t_2/t_1=(2-\sqrt{2})/2$ \cite{Sun} -- here, however, we omit this 
term because we are only interested in the Berry curvature, which does not depend on $d_0(\bk)\bone$, as mentioned above.

The Berry curvature of the checkerboard-lattice model, calculated from Eq. (\ref{eq12}) is shown in Fig. \ref{fig02}, for the 
choice of parameters $t_2/t_1=(2-\sqrt{2})/2$, $\phi=\pi/4$, and $m/t_1=1$. In this case, the gap between the two bands is minimal at
the point $\bk_Da_0=(\pi,0)$ at the border of the first Brillouin zone, $\Delta=m-4t_2\simeq -0.17 t_1$, and the Berry curvature is strongly
peaked there [Fig. \ref{fig02}(a)]. The figures \ref{fig02}(b) and (c) show cuts through the Berry curvature for $k_y=0$ and $k_x=\pi$,
respectively (full line) in comparison with the Dirac-point approximation (dashed lines) given by Eq. (\ref{eq14}), 
where the effective parameters are related to those of the lattice model by
\beqn
M &=& m-4t_2,\\
v_x &=& 2t_1 a_0 \cos(\phi),\\
v_y &=& 2t_1 a_0 \sin(\phi),
\eeqn
as one easily obtains from a series expansion around the point $\bk_D$. One notices that the Dirac-point approximation yields an 
excellent agreement with the full Berry curvature, as one may expect for a strongly peaked Berry curvature that is zero in the major
part of the first Brillouin zone and that does not fulfil
the criterion (\ref{eq15}), since $\kappa_{x/y}=M/v_{x/y}\simeq 0.12/a_0$, i.e. close to a topological phase transition. Indeed,
the topological phase transition to a band insulator with zero Chern number occurs when $M$ changes its sign, that is for 
$m=4t_2\simeq 1.17t_1$, where the band dispersion reveals true (gapless) Dirac points at $\bk_D$ and where the Berry curvature is reduced
to delta functions around this point, before it changes sign and vanishes completely upon further increase of $m$.

\begin{figure}
\centering
\includegraphics[width=5.5cm,angle=0]{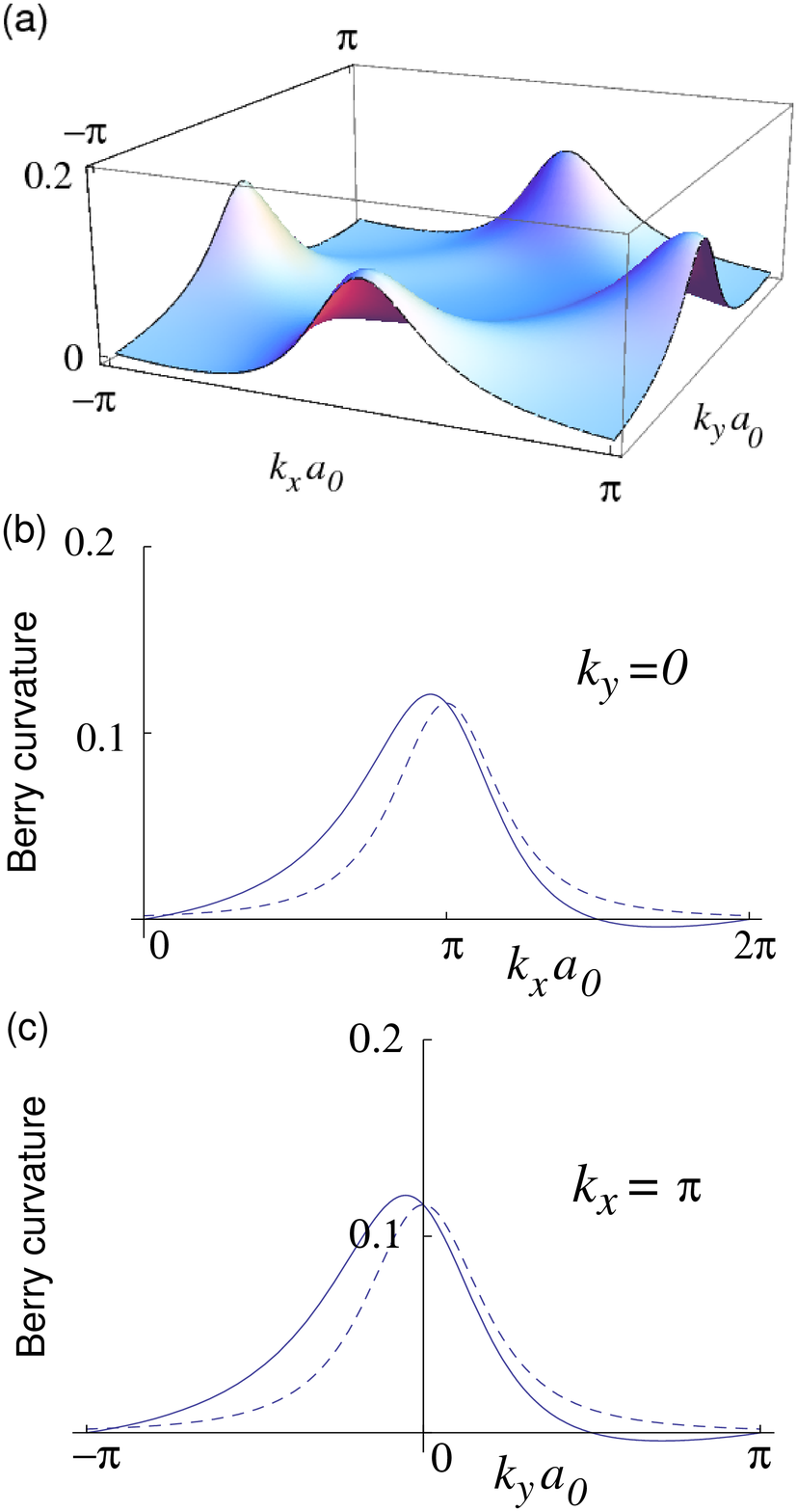}
\caption{\footnotesize{(a) Berry curvature of the checkerboard-lattice model with $t_2/t_1=(2-\sqrt{2})/2$, $\phi=\pi/4$, and $m/t_1=0$, 
over the whole Brillouin zone. (b) Berry curvature for $k_y=0$ (full line), in comparison to approximate formula (\ref{eq14}) (dashed
line). (c) Same for $k_x=\pi$.
}}
\label{fig03}
\end{figure}

The situation changes when decreasing the parameter $m$, in which case the band gap that is now dominated by time-reversal-symmetry 
breaking is increased. In fact, for $m=0$,
there is a second (avoided) Dirac point present at $\bk_D^{\prime} a_0=(0,\pi)$, where the band gap $\Delta=M=-4t_2\simeq -1.17t_1$
is identical to that at 
$\bk_D$. The Berry curvature for the same parameter choice as for Fig. \ref{fig02}, 
but with $m=0$ is shown in Fig. \ref{fig03}. Notice that the criterion
(\ref{eq15}) for the flatness of the Berry curvature is now better fulfilled since $\kappa_{x/y}a_0\simeq 0.83$, in agreement with
Fig. \ref{fig03}(a), where the change of the vertical scale for the Berry curvature should be emphasised when compared to 
Fig. \ref{fig02}(a). Figures \ref{fig03}(b) and (c) show a cut of the Berry curvature at $k_y=0$ and $k_x=0$, respectively. As a 
consequence of the large value of $\kappa_{x/y}$, the contributions to the curvature emanating from the two different avoided Dirac points
$\bk_D$ and $\bk_D^{\prime}$ now overlap, such that the Berry curvature is effectively flattened and the
Dirac-point approximation yields only a semi-quantitative agreement. Whereas
it provides a satisfactory agreement in the maximal height of the Berry curvature, it does not account for the anisotropy around the
avoided Dirac points. However, also the order of magnitude of the peak width is correctly accounted for in the Dirac-point approximation
(\ref{eq14}).

%\subsection{Single-band wave functions}

%\subsection{Pseudopotentials}

\section{Fractional Quantum Spin Hall Effect}

% general algebraic properties : total density is normal, relative density; how to incorporate cross correlations between different components

The quantum spin Hall effect arises in time-reversal symmetric systems, e.g. in the presence of a strong spin-orbit coupling \cite{TIrev}.
In this case, the two partners $\sigma=\ua,\da$ of a Kramers pair, such as the two spins,\footnote{For simplicity, we consider spin 
in the following paragraphs to denote the two partners of a Kramers pair, but the approach remains valid if the Hamiltonian
is block-diagonal in a generic $\mathbbm{Z}_2$ quantum number denoted by $\sigma=\ua,\da$.} 
may be viewed each as a Chern insulator, and the one-particle energy may be described by the Hamiltonian
\beq
H_{QSHE}(\bk)=\left(
\begin{array}{cc} H^{\ua}(\bk) & 0 \\ 0 &H^{\da}(\bk) \end{array}
\right),
\eeq
where $H^{\sigma}(\bk)=H^{-\sigma}(-\bk)^*$ is the Hamiltonian for a Chern insulator band of spin-$\sigma$ electrons.
Time-reversal symmetry imposes that the Berry curvature of one spin orientation is related to the other one by
\beq\label{eq16}
\Bmath_{\lambda,\bk;\sigma}=-\Bmath_{\lambda,-\bk;-\sigma},
\eeq
within the same orbital band $\lambda$.

\subsection{Algebraic properties in the flat-curvature limit}

If we consider, as in the previous section, a homogeneous Berry curvature in reciprocal space, the same reasoning as that 
presented in the context of fractional Chern insulators and Eq. (\ref{eq16}) indicate that
spin-$\ua$ electrons experience a Berry curvature $\Bmath_{\lambda,\ua}$ 
with an opposite sign as that $\Bmath_{\lambda,\da}$ of spin-$\da$ electrons, and we use the notation 
$\Bmath_{\lambda\sigma}=\sigma\Bmath_{\lambda}$ from now on (with $\sigma=+$ for spin-$\ua$ and $-$ for spin-$\da$ electrons).
The GMP algebra therefore becomes spin-dependent and reads
\beq\label{eq17}
\left[\rhobar_{\lambda,\sigma}(\bq),\rhobar_{\lambda,\sigma'}(\bq')\right]= 
2i\sigma \sin\left(\frac{\bq\wedge\bq'}{2}\Bmath_{\lambda}\right)\rhobar_{\lambda,\sigma}(\bq+\bq')\delta_{\sigma,\sigma'}.
\eeq
One may furthermore introduce the total density 
$\rhobar_{tot,\lambda}(\bq)=\rhobar_{\lambda,\ua}(\bq)+\rhobar_{\lambda,\da}(\bq)$ and the Fourier component of the local spin polarisation 
in the $z$-direction $\Sbar_{\lambda}^z(\bq)=\rhobar_{\lambda,\ua}(\bq)-\rhobar_{\lambda,\da}(\bq)$. The algebra for these operators is 
then given by the commutation relations
\beqn\label{eq18}
\nn
\left[\rhobar_{tot,\lambda}(\bq),\rhobar_{tot,\lambda}(\bq')\right] &=& 
2i \sin\left(\frac{\bq\wedge\bq'}{2}\Bmath_{\lambda}\right)\Sbar_{\lambda}^z(\bq+\bq'),\\
\nn
\left[\Sbar_{\lambda}^z(\bq),\Sbar_{\lambda}^z(\bq')\right] &=& 
2i \sin\left(\frac{\bq\wedge\bq'}{2}\Bmath_{\lambda}\right)\Sbar_{\lambda}^z(\bq+\bq'),\\
\nn
\left[\Sbar_{\lambda}^z(\bq),\rhobar_{tot,\lambda}(\bq')\right] &=& 
2i \sin\left(\frac{\bq\wedge\bq'}{2}\Bmath_{\lambda}\right)\rhobar_{tot,\lambda}(\bq+\bq'),\\
\eeqn
which is different from the SU(2) extension of the GMP algebra, $W_{\infty}(2)$ \cite{ezawa}, which describes 
the dynamics of two-component electrons in conventional Landau levels. This difference arises from the fact that
spin-$\ua$ and spin-$\da$ electrons experience an effective magnetic field with a different sign. Indeed, the
first commutation relation in Eq. (\ref{eq18}) shows that, in the absence of a spin polarisation ($\langle \Sbar_{\lambda}^z \rangle=0$),
the total electronic density behaves on the mean-field level as if there was no Berry curvature at all. The algebra (\ref{eq18})
thus translates the fact that the total Chern number, and thus the Hall conductivity, vanishes in the quantum spin Hall 
effect \cite{TIrev}.

\subsection{Wave functions}

In the lines of the discussion of the partially filled Chern band, we discuss the FQSHE in terms of wave functions in the 2D plane. 
The starting point in the construction of single-particle wave functions is Eq. (\ref{eq04}) for a flat Berry curvature, in which 
case we may use the symmetric gauge 
\beq\label{eq19}
\Amath_{\lambda,\sigma}(\bk)=\frac{\Bmath_{\lambda,\sigma}}{2}\bk\times\be_z = \frac{\sigma\Bmath_{\lambda}}{2}\bk\times\be_z.
\eeq
The spin-dependent guiding-centre operator, which then reads (we omit the hats on the position operators from now on)
\beq\label{eq20}
\bR_{\lambda,\sigma}\equiv \left(X_{\sigma},Y_{\sigma}\right)=\br - \frac{\sigma\Bmath_{\lambda}}{2}\bk\times\be_z,
\eeq
satisfies the commutation relation
%\beq\label{eq21}
$\left[X_{\sigma},Y_{\sigma'}\right]= -i\sigma\Bmath_{\lambda}\delta_{\sigma,\sigma'},$
%\eeq
which agrees naturally with the spin-dependent GMP algebra (\ref{eq17}) if one identifies 
$\rhobar(\bq)=\exp(i\bq\cdot\bR_{\lambda,\sigma})$ in the one-particle description. In this case, one may introduce harmonic-oscillator
ladder operators
\beq\label{eq22}
b_{\sigma}=\frac{1}{\sqrt{2\Bmath_{\lambda}}}\left(X_{\sigma} - i\sigma Y_{\sigma}\right), ~
b_{\sigma}^{\dagger}=\frac{1}{\sqrt{2\Bmath_{\lambda}}}\left(X_{\sigma} + i\sigma Y_{\sigma}\right)
\eeq
that satisfy the usual commutation relations $[b_{\sigma},b_{\sigma'}^{\dagger}]=\delta_{\sigma,\sigma'}$. 
These ladder operators may then be represented in terms of operators acting on (anti-)analytic functions,
\beqn\label{eq23}
\nn
&& b_{\ua}\sim \sqrt{\Bmath_{\lambda}}\partial_{z_{\ua}}, \qquad b_{\ua}^{\dagger} \sim z_{\ua}/\sqrt{\Bmath_{\lambda}} \\
&& b_{\da}\sim \sqrt{\Bmath_{\lambda}}\partial_{z_{\da}^*}, \qquad b_{\da}^{\dagger} \sim z_{\da}^*/\sqrt{\Bmath_{\lambda}},
\eeqn
such that the eigenstates within the band $\lambda$ may be labeled by the quantum number $m_{\sigma}$ associated with the operator
$b_{\sigma}^{\dagger}b_{\sigma}$. Here, the star indicates complex conjugation, i.e. $z^*=(X-iY)/\sqrt{2}$ is the complex 
conjugate of $z=(X+iY)/\sqrt{2}$.
One thus obtains (anti-)analytic one-particle wave functions similarly to the Landau-level problem
of electrons in a quantising magnetic field,
\beq\label{eq24}
%\nn
\phi_{m_{\ua}}^{\ua}(z_{\ua}) \propto \left(z_{\ua}/\sqrt{\Bmath_{\lambda}}\right)^{m_{\ua}},\qquad
\phi_{m_{\da}}^{\da}(z_{\da}^*) \propto \left(z_{\da}^*/\sqrt{\Bmath_{\lambda}}\right)^{m_{\da}}.
\eeq
Notice that the difference in the sign of the Berry curvature for the two spin orientations leads to analytic 
wave functions for one spin orientation ($\ua$), whereas those for the other orientation ($\da$) are anti-analytic. 

In the case of a correlated $N$-particle state, such as the FQSHE, one is therefore confronted with hybrid wave functions,
\beqn\label{eq25}
\nn
\Psi(\{z_{k,\ua},z_{l,\da}^*\}) &=& \prod_{k<l}\left(z_{k,\ua} -z_{l,\ua}\right)^{\mtilde_{\ua}}\prod_{k<l}\left(z_{k,\da}^* -z_{l,\da}^*\right)^{\mtilde_{\da}}\\
&& \times \chi(\{z_{k,\ua},z_{l,\da}^*\}),
\eeqn
which consist of a product of two Jastrow factors,
where $\mtilde_{\sigma}$ plays the role of the relative angular momentum between particles with spin orientation $\sigma$.
As a consequence of fermion statistics, the exponents $\mtilde_{\sigma}$ must be odd integers. The wave functions (\ref{eq25}) 
reflect a class of states that has been investigated before in the framework of topological Chern-Simons-type theories \cite{LS,CM}.
The additional term $\chi(\{z_{k,\ua},z_{l,\da}^*\})$ takes into account possible correlations between spin-$\ua$ and spin-$\da$
electrons. However, in contrast to two-component quantum Hall systems, such as a quantum Hall bilayer or simply one that takes into
account the electronic spin, this term is not expected to be the conventional Jastrow factor of Halperin's wave 
function \cite{halperin}, because of the mixed analyticity. This is also stipulated by the algebraic structure (\ref{eq18}) that
is different from the SU(2) extension of the GMP algebra that describes two-component quantum Hall systems.

To illustrate this point, we consider the Jastrow-type factor $(z_{k,\ua}-z_{l,\da}^*)^n$ that would occur in Halperin's wave 
function accounting for possible correlations between electrons of different spin orientation and that has been considered
by Bernevig and Zhang in a continuum description of the FQSHE \cite{BZ06}. This factor enforces a zero for 
$X_{k,\ua}=X_{l,\da}$ and $Y_{k,\ua}=-Y_{l,\da}$. The zero in the wave function would be useful, as in the case of the usual
Laughlin wave function \cite{laughlin}, if the interaction potential is strongly repulsive at short distances between the 
particle positions, that is $z_{k,\ua}\sim z_{l,\da}$. Therefore, the zeros of the Jastrow factor do not coincide with the short
distances, and the zeros screen roughly an interaction strength $V(r\sim 2|Y_{\ua}|)$ rather than $V(r\sim 0)\gg V(r\sim 2|Y_{\ua}|)$.
The above-mentioned Jastrow factor is therefore much less efficient than the corresponding factor in usual two-component wave functions
such as Halperin's. Furthermore, a two-particle wave function for electrons with opposite spin may generally be written
as $\psi_{M,m}\sim Z^M z^m$, with $z=z_{1,\ua}-z_{2,\da}^*$ and $Z=z_{1,\ua}+z_{2,\da}^*$. However, this two-particle wave function
is not an eigenstate of an interaction potential $V(|z_{1,\ua}-z_{2,\da}|)$ that depends only on the distance between the 
electron coordinates, such that a pseudo-potential expansion \cite{PP} would not capture inter-component correlations.

Based on these arguments, we therefore briefly comment on the special case where we neglect inter-component correlations, that 
is $\chi(\{z_{k,\ua},z_{l,\da}^*\})=1$. The wave functions (\ref{eq25}) describe then a possible FQSHE at filling
factors 
\beq
\nu=2\pi \Bmath_{\lambda} n_{el}=\frac{1}{\mtilde_{\ua}}+\frac{1}{\mtilde_{\da}},
\eeq 
and the FQSHE may then be viewed as a product of two
(uncorrelated) Laughlin states, one for each spin orientation. The stability of the state with $\mtilde_{\ua}=\mtilde_{\da}=3$ 
has recently been investigated numerically by Neupert \textit{et al.}, who found that inter-component correlations may destroy
such a state \cite{Neupert2}. Whether other such FQSHE states may be stabilised, as well as to what 
extent the mapping to the 2D plane describes accurately in the thermodynamic limit the compact geometry of the first Brillouin zone, 
might be answered by further numerical investigations. Also the true form of the inter-component correlations 
$\chi(\{z_{k,\ua},z_{l,\da}^*\})$ remains an open issue for future theoretical studies.

\section{Conclusions}

To summarise, we have analysed the algebraic structure of the FQAHE in Chern insulators and that of a possible FQSHE. These effects
arise naturally in infinitely flat energy bands with a flat (non-zero) Berry curvature. In this case, the 
global topological properties of the energy bands, that is the non-zero Chern numbers associated with the band in the case of the 
FQAHE or those for the two Kramers partners in the case of the FQSHE, may be tracked down to local properties encoded in the GMP 
algebra of the projected electronic density operators in Fourier space. Alternatively, this algebra may be viewed as arising from
a local Aharonov-Bohm effect that yields a non-commutative geometry. Whereas the criterion for the flatness of the energy bands is
rather straight-forward in the sense that the band dispersion must be small as compared to the gaps arising in the formation of
incompressible FQHE liquids, the criterion for the flatness of the Berry curvature is more involved. It invokes the underlying 
(avoided) Dirac-point structure the mass gap of which yields a characteristic wave vector below which the Berry curvature may 
be viewed as approximately constant. Once this characteristic wave vector is on the order of the size of the first Brillouin zone,
$\kappa_{x/y}\sim 1/a_0$, the Berry curvature may be approximated by a constant value. 

When compared to the FQHE in 2D electronic
systems in a strong magnetic field, the flat-band assumption may be identified with the limit of vanishing disorder, whereas the
flat-curvature approximation corresponds to a homogeneous distribution of the flux density in the 2D plane. Because of the energy
gap, which accompanies the formation of incompressible quantum liquids in the FQHE, small deviations from the flat-band and 
flat-curvature limits do not alter the overall picture of the FQAHE and the FQSHE. In the case of the FQSHE, the spin orientation
of the electrons determines whether they are described in terms of analytic or anti-analytic wave functions. This situation
is strikingly different from that of two-component quantum Hall systems that may be described, on the algebraic level, in terms of
an SU(2) extension of the GMP algebra [or $W_{\infty}(2)$] or in terms of Halperin wave functions. 

\subsection*{Acknowledgments}

The author thanks J.-N. Fuchs, F. D. M. Haldane, F. Pi\'echon, and N. Regnault for fruitful discussions.


\begin{thebibliography}{99}

\bibitem{TSG}D.\ C.\ Tsui, H.\ Stormer, and A.\ C.\ Gossard, Phys.\ Rev.\ Lett.\ {\bf 48}, 1559 (1982).

\bibitem{laughlin}R.\ B.\ Laughlin, Phys.\ Rev.\ Lett.\ {\bf 50}, 1395 (1983).

\bibitem{TKNN} D. J. Thouless, M. Kohmoto, M. P. Nightingale, and M. den Nijs, Phys. Rev. Lett. {\bf 49}, 405 (1982).

\bibitem{TIrev}For a review, see M. Z. Hasan and C. L. Kane, Rev. Mod. Phys. {\bf 82}, 3045 (2010); 
X.-L. Qi and S. C. Zhang, Rev. Mod. Phys. {\bf 83}, 1057 (2011).

\bibitem{haldane88} F. D. M. Haldane, Phys. Rev. Lett. {\bf 61}, 2015 (1988).

\bibitem{KM05}C. L. Kane and E. J. Mele, Phys. Rev. Lett. {\bf 95}, 226801 (2005).

\bibitem{BHZ}B. A. Bernevig, T. A. Hughes, and S. C. Zhang, Science {\bf 314}, 1757 (2005).

\bibitem{molenkamp}M. K\"onig, S. Wiedmann, C. Br\"une, A. Roth, H. Buhmann, L. W. Molenkamp, X. L. Qi, and S. C. Zhang,
Science {\bf 318}, 766 (2005).

\bibitem{Tang}E. Tang, J.-W. Mei, and X.-G. Wen, Phys. Rev. Lett. {\bf 106}, 236802 (2011).

\bibitem{Neupert1}T. Neupert, L. Santos, C. Chamon, and C. Mudry, Rev. Lett. {\bf 106}, 236804 (2011).

\bibitem{Sun}K. Sun, Z.-C. Gu, H. Katsura, and S. Das Sarma, Rev. Lett. {\bf 106}, 236803 (2011).

\bibitem{Fiete}X. Hu, M. Kargarian, and G. A. Fiete, Phys. Rev. B {\bf 84}, 155116 (2011).

\bibitem{Sheng}Y.-F. Wang, Z.-C. Gu, K. Sun, C.-D. Gong, and D. N. Sheng, Phys. Rev. Lett. {\bf 107}, 146803 (2011).

\bibitem{RB}N. Regnault and B. A. Bernevig, Phys. Rev. X {\bf 1}, 021014 (2011).

\bibitem{Qi}X.-L. Qi, Phys. Rev. Lett. {\bf 107}, 126803 (2011).

\bibitem{PRS}S. A. Parameswaran, R. Roy, and S. L. Sondhi, arXiv:1106.4025.

\bibitem{GMP}S.\ M.\ Girvin, A.\ H.\ MacDonald, and P.\ M.\ Platzman, Phys.\ Rev.\ B\ {\bf 33}, 2481 (1986).

\bibitem{Neupert2}T. Neupert, L. Santos, S. Ryu, C. Chamon, and C. Mudry, Phys. Rev. B {\bf 84}, 165107 (2011).

\bibitem{PP} F. D. M. Haldane, Phys. Rev. Lett. {\bf 51}, 605 (1983).

\bibitem{inhomFQHE}A. L. Efros, Phys. Rev. B {\bf 45}, 11\,354 (1992).

\bibitem{fuchs}J.-N. Fuchs, F. Pi\'echon, M. O. Goerbig, and G. Montambaux,
Eur. Phys. J. B {\bf 77}, 351 (2010).

\bibitem{ezawa}Z. F. Ezawa, G. Tsitsishvili, and K. Hasebe, Phys. Rev. B {\bf 67}, 125314 (2003).

\bibitem{LS} M. Levin and A. Stern, Phys. Rev. Lett. {\bf 103}, 196803 (2009).

\bibitem{CM} G. Y. Cho and J. E. Moore, Annals Phys. {\bf 326} 1515 (2011).

\bibitem{halperin}
B. I. Halperin, Helv. Phys. Acta {\bf 56}, 75 (1983).

\bibitem{BZ06} B. A. Bernevig and S.C. Zhang, Phys. Rev. Lett. {\bf 96}, 106802 (2006).


\end{thebibliography}
\end{document}